\documentclass{iaus}
\usepackage{graphicx}

\title[JD 11.~~The oblate finite disk correction factor] 
{Radiation driven winds with rotation:\\ The oblate finite disc correction factor}

\author[I. Araya, M. Cur\'e, A. Granada \& L. Cidale]   
{Ignacio Araya,$^1$ Michel Cur\'e,$^1$ Anah\'i Granada$^2$ \& Lydia Cidale$^3$}

\affiliation{$^1$Departamento de F\'isica y Astronom\'ia, Facultad de Ciencias, Universidad de Valpara\'iso, Chile\\
$^2$Facultad de Ciencias Astron\'omicas y Geof\'{\i}sicas, Universidad Nacional de La Plata, Argentina\\
$^3$Instituto de Astrof\'{\i}sica La Plata, CCT La Plata-CONICET-UNLP, Argentina\\
$^4$Observatoire de Geneve. Universit\'e de Geneve, Suisse\\[\affilskip]}

\pubyear{2010}
\volume{272}  
\pagerange{119--126}
\jname{Active OB stars: structure, evolution, mass loss and critical limits}
\editors{A.C. Editor, B.D. Editor \& C.E. Editor, eds.}

\begin{document}
\maketitle
\begin{abstract}

We have incorporated the oblate distortion of the shape of the star due to the 
stellar rotation, which modifies the finite disk correction factor ($f_{D}$) in the m-CAK 
hydrodynamical model. We implement a simplified version for the $f_{D}$ allowing us 
to solve numerically the non--linear m-CAK momentum equation. We solve this model for a 
classical Be star in the polar and equatorial directions. The star's oblateness modifies the 
polar wind, which is now much faster than the spherical one, mainly because the wind receives 
radiation from a larger (than the spherical) stellar surface. In the equatorial direction we obtain 
slow solutions, which are even slower and denser than the spherical ones. For the case when the 
stellar rotational velocity is about the critical velocity, the most remarkable result of our 
calculations is that the density contrast between the equatorial density and the polar one, 
is about 100. This result could explain a long-standing problem on Be stars.

\end{abstract}

\firstsection 
\section{Introduction}
Pelupessy et al. (2000) formulated the wind momentum equation for sectorial line driven winds
including the finite disk correction factor for an oblate rotating star with gravity darkening for
both the continuum and the lines. They calculated models with line--force parameters around
the bi--stability jump at 25\,000 K. In this case, from the pole to the equator, the mass flux
increases and the terminal velocity decreases. Their results showed a wind density contrast
$ \rho (equator) /  \rho (pole)$ (hereafter $\rho_{e} /  \rho_{p}$)  of about a factor 10 independent
of the rotation rate of the star.

In this work, we implement an approximative version for the oblate finite--disk correction factor, $f_{O}$,
allowing us to solve numerically the non linear m--CAK momentum. We solve then  this equation for
a {\it{classical Be}} star, for polar and equatorial directions. In this study we do not take into account the
bi--stability jump.

\section{Oblate Factor}
In order to  incorporate the oblate distortion of the shape of the star due to the stellar rotation to the
m--CAK hydrodynamic model, we implement an approximative function.  In view of the behaviour
of $f_{O}$ and $f_{D}$ we approximate its ratio $f_{O}/f_{D}$ via a sixth order polynomial interpolation 
in the inverse radial variable $u=-R_{\star}/r$, i.e., $f_{O}=Q(u)\,f_{D}$.
In this form, we assure that the topology found by Cur\'e (2004) is maintained by the $f_{D}$ term, but it is
modified by the incorporation of the $Q(u)$ polynomial. With this approximation we can solve numerically
the non--linear m--CAK differential equation.

\section{Results}
 We solve the oblate m--CAK equations for a classical Be star with the following stellar parameters:
$T_{eff}=25\,000$ K, $log\, g=4.03$, and $R/R_{\odot}=5.3$ (Slettebak et al. 1980) and line--force
parameter $k=0.3$. For the other parameters, we have used two different values for $\delta$,
namely:  $\delta=0.07$ and $\delta=0.15$, and three values for $\alpha$, i.e., $\alpha=0.4, 0.45, 0.55$.
In this study, we have considered  {\it{the same value of $\alpha$, $k$ and $\delta$}} for the pole
and equator, i.e.  without taking into account the bi--stability jump. We show only solutions for
$\omega=0.99$ at the equator and pole. For the cases where $\alpha=0.4$ and $\alpha=0.45$, the density
contrasts exceed a factor 100, values which are in agreement with observations  (Lamers \&  Waters 1987).

\begin{figure}[h]
\begin{center}
 \includegraphics[width=2.6in]{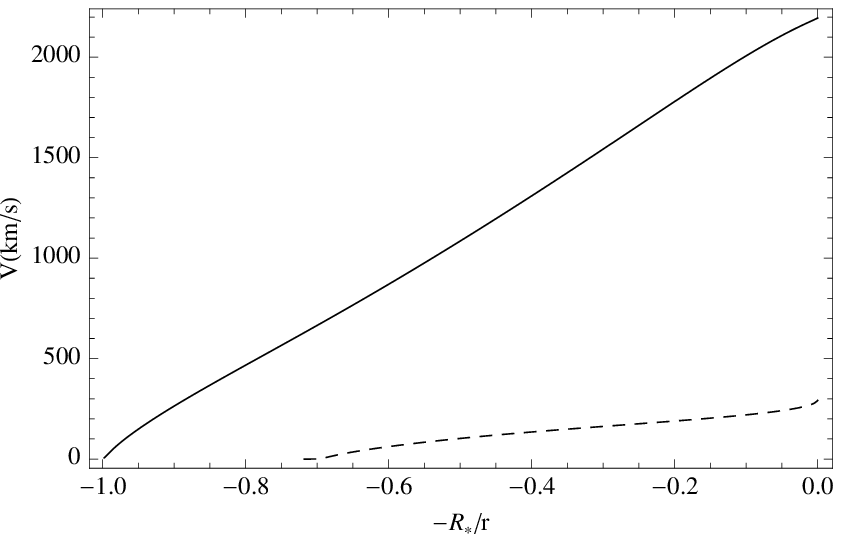}
 \includegraphics[width=2.6in]{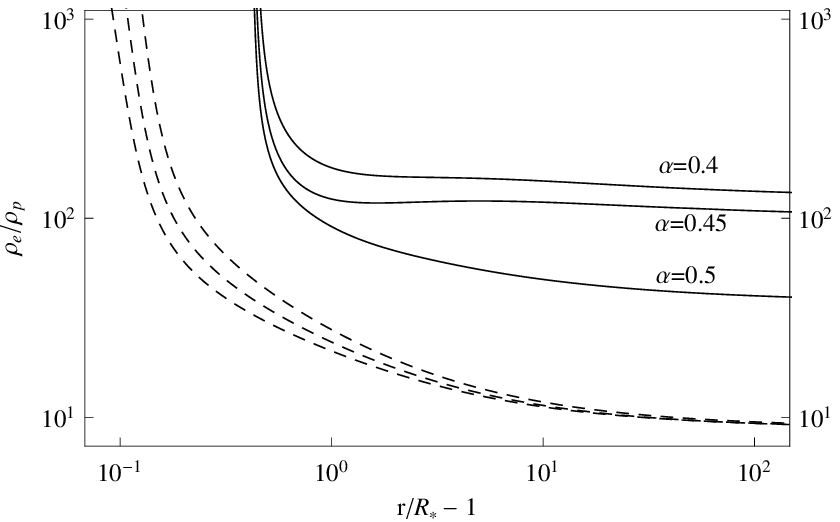}\\
 Fig. 1a \hspace{2 in}  Fig. 1b
\caption{a) Velocity versus $u$ for equator (\textit{dashed line}) and pole (\textit{solid line}) 
with $\alpha=0.4$ and $\delta=0.07$.  b) Solid lines: Density constrasts for 
$\omega=0.99$ and  $\delta=0.07$. Dashed lines: Spherical cases.}
   \label{fig1}
\end{center}
\end{figure}

\section{Summary and Conclusions}

The oblate correction factor has been implemented in an approximative form. The factor $Q_{r}$ in the
oblate correction factor certainly modifies the topology of the hydrodynamical differential equation and we suspect from
first calculations that other critical points may exists and, therefore, more solutions might be present. We recover the
observed density contrast only when the rotational velocity of the star is near the break-up velocity,
confirming other theoretical works (see e.g., Townsend et al, 2004). The use of a set of self--consistent
line force parameters is necessary to understand the wind dynamics of these rapid rotators. The full 
version of the oblate correction factor will be implemented to study the topology  of the wind.

\end{document}